\documentclass[usenatbib]{mnras}

\usepackage{graphicx}
\usepackage{amsmath}
\usepackage{amssymb}

\usepackage{orcidlink}
\usepackage{csquotes}
\usepackage{listings}
\usepackage{array}
\usepackage{xcolor}
\usepackage{enumitem}
\setlist[itemize]{left=0pt, labelsep=1em, label=\textbullet, itemsep = 0.4em}
\usepackage{enumitem}
\setlist[enumerate,1]{label=\arabic*.}
\setlist[enumerate]{topsep=0.6em, labelsep=1em, itemsep = 0.6em}
\definecolor{mygreen}{RGB}{41, 150, 70}
\definecolor{red}{RGB}{191, 38, 27}
\definecolor{orange}{RGB}{245, 137, 2}
\usepackage{subcaption}
\usepackage{soul}

\usepackage{graphicx}
\usepackage{longtable}

\def\lesssim{\mathrel{\hbox{\rlap{\hbox{\lower4pt\hbox{$\sim$}}}\hbox{$<$}}}}
\def\gtrsim{\mathrel{\hbox{\rlap{\hbox{\lower4pt\hbox{$\sim$}}}\hbox{$>$}}}}

\newcommand{\msun}{\mbox{M$_{\odot}$}}
\newcommand{\msol}{\mbox{M$_{\odot}$}}

\newcommand{\kms}{\mbox{$\rm{km}\,s^{-1}$}}

\newcommand{\Ni}{\mbox{$^{56}$Ni}}

\newcommand{\ergs}{erg\,s$^{-1}$}

\newcommand{\cow}{\mbox{AT\,2018cow}}
\newcommand{\qfm}{\mbox{AT\,2024qfm}}

\title[AT\,2024qfm: a luminous fast blue optical transient]
{AT\,2024qfm: a luminous fast blue optical transient at a redshift of $z = 0.2267$ identified by Lasair-ZTF}

\author[M.~Fulton et al.]{
    M.~Fulton$^{1}$\thanks{E-mail: m.fulton@qub.ac.uk}\orcidlink{0000-0003-1916-0664}, 
    S.~J.~Smartt$^{2,1}$\orcidlink{0000-0002-8229-1731},
    S.~Srivastav$^2$\orcidlink{0000-0003-4524-6883},  
    J.~H.~Gillanders$^2$\orcidlink{0000-0002-8094-6108},
    J.~W.~Tweddle$^{2}$\orcidlink{0009-0004-5681-545X},
    \newauthor
    M.~E.~Huber$^{3}$\orcidlink{0000-0003-1059-9603},
    M.~Nicholl$^{1}$\orcidlink{0000-0002-2555-3192}, 
    C.~R.~Angus$^{1}$\orcidlink{0000-0002-4269-7999},
    K.~W.~Smith$^{2,1}$\orcidlink{0000-0001-9535-3199}, 
    K.~C.~Chambers$^{3}$\orcidlink{0000-0001-6965-7789},
    \newauthor
    A.~Lawrence$^{4}$\orcidlink{0000-0002-9573-5637},
    R.~Williams$^{4}$\orcidlink{0009-0006-9214-4520},
    D.~R~Young$^{1}$\orcidlink{0000-0002-1229-2499},   
    K.~Auchettl$^{5}$\orcidlink{0000-0002-4449-9152},
    T.~de~Boer$^{3}$\orcidlink{0000-0001-5486-2747}, 
    \newauthor
    T.-W.~Chen$^{6}$\orcidlink{0000-0002-1066-6098},
    C.-H.~Lai$^{6}$\orcidlink{0009-0006-7037-0594},
    C.~C.~Lin$^{3}$\orcidlink{0000-0002-7272-5129},
    G.~S.~H.~Paek$^{3}$\orcidlink{0000-0002-6639-6533}, 
    M.~Pursiainen$^{7}$\orcidlink{0000-0003-4663-4300},
    \newauthor
    S.~I.~Raimundo$^{8}$\orcidlink{0000-0002-6248-398X}, 
    R.~Wainscoat$^{3}$\orcidlink{0000-0002-1341-0952} and
    S.~Yang$^{9}$\orcidlink{0000-0002-2898-6532}
    \\
    $^{1}$Astrophysics Research Centre, School of Mathematics and Physics, Queen's University Belfast, BT7 1NN, UK\\
    $^{2}$Astrophysics sub-Department, Department of Physics, University of Oxford, Keble Road, Oxford, OX1 3RH, UK\\
    $^{3}$Institute for Astronomy, University of Hawai'i, 2680 Woodlawn Drive, Honolulu, HI 96822, USA\\
    $^{4}$Institute for Astronomy, University of Edinburgh, Royal Observatory, Blackford Hill, Edinburgh EH9 3HJ, UK\\  
    $^{5}$OzGrav, School of Physics, The University of Melbourne, Parkville, VIC 3010, Australia\\
    $^{6}$Graduate Institute of Astronomy, National Central University, 300 Jhongda Road, 32001 Jhongli, Taiwan\\
    $^{7}$Department of Physics, University of Warwick, Gibbet Hill Road, Coventry CV4 7AL, UK\\
    $^{8}$School of Physics and Astronomy, University of Southampton, Highfield, Southampton, SO17 1BJ, UK\\
    $^{9}$Institute for Gravitational Wave Astronomy, Henan Academy of Sciences, Zhengzhou 450046, Henan, China
}

\pubyear{2026}

\begin{document}
\label{firstpage}
\pagerange{\pageref{firstpage}--\pageref{lastpage}}
\maketitle

\begin{abstract}
    Luminous fast blue optical transients (LFBOTs) emit from x-ray to radio wavelengths, epitomised by the discovery of AT\,2018cow in a host galaxy at 65\,Mpc. In the following eight years eleven more have been found, at redshifts $0.075 \lesssim z \lesssim0.34$, plus one identified retrospectively from 2016. Here we present the discovery of AT\,2024qfm, classified as an LFBOT in a host galaxy at $z = 0.2267 \pm 0.0002$. Its ultraviolet-to-optical luminosity and rapid 13~day fade closely match AT\,2018cow. We describe how the transient was identified in the Zwicky Transient Facility alert stream using a custom filter in the Lasair broker that flags flux gradients over time. Another LFBOT candidate was identified with the same methodology (AT\,2024kth). The physical origin of LFBOTs remains debated with no firm consensus, and further progress requires more discoveries, host-galaxy characterisation, and multi-wavelength analysis to constrain theory. We discuss this discovery in the context of Rubin Observatory's Legacy Survey of Space and Time (LSST), whose sensitivity will increase the effective LFBOT survey volume tenfold relative to ZTF, out to $z \lesssim 0.6$, and show that our FastFinder filter could recover such events. We highlight the challenge of detecting their fast evolution with sufficiently low latency to trigger multi-wavelength follow-up that can constrain theoretical models.
\end{abstract}


\section{Introduction}
\label{sec:intro}

The era of wide-field optical surveys that have the ability to survey the whole visible sky with a $1 - 2$~day cadence has revealed fast-evolving, extragalactic transients. These have been shown to be multi-wavelength emitters, from X-ray to the radio regime, indicating non-thermal processes and likely relativistic jets. These luminous fast blue optical transients (LFBOTs) and are characterised by rapid photometric evolution along with persistently blue, hot, and featureless spectra. With peak blue magnitudes well above $M_g\sim-20$ and bolometric luminosities $L_{\rm bol} > 10^{44}$\,\ergs, their physical nature is difficult to reconcile with standard core-collapse mechanisms, thermonuclear supernovae or \Ni\ powering. 

The prototype of this class of objects is \cow, discovered by the Asteroid Terrestrial-impact Last Alert System \citep[ATLAS;][]{Tonry2018_ATLAS, Smith2020_ATLAS} with an extreme rise rate and peak luminosity \citep{2018ApJ...865L...3P,Smartt2018}. Multi-wavelength follow-up established a peak optical luminosity approaching that of superluminous supernovae, a near-featureless blue continuum at early times, luminous and long-lived radio and millimetre emission from a mildly relativistic outflow, and highly variable, luminous X-ray emission indicative of a compact central engine \citep{Rivera2018,Ho2019,2019ApJ...872...18M,Perley2019,Kuin2019}. The extreme and multi-wavelength nature of \cow\ immediately distinguished it from the broader, more common population of `fast blue optical transients' (FBOTs), most of which are now understood to be core-collapse events with low ejecta mass \citep{Drout2014,Pursiainen2018}. 

In the years since, systematic searches of Zwicky Transient Facility \citep[ZTF;][]{Bellm2019, 2019PASP..131a8001P} alert-stream data, informed by the fast rise, high peak luminosity, and blue colour of \cow, have uncovered a  growing number of further examples: the Koala (AT\,2018lug/ZTF18abvkwla) at $z=0.271$, hosted by a dwarf  galaxy with radio emission an order of magnitude more luminous than \cow\ \citep{Ho2020}; CSS161010, a nearby ($z=0.033$) event with a mildly relativistic radio-emitting outflow located in an extremely low-mass dwarf host \citep{Coppejans2020}; the Camel (AT\,2020xnd/ZTF20acigmel) at $z=0.243$, confirmed as an \cow\ analogue in real time \citep{Perley2021}; the X-ray and radio-loud AT\,2020mrf at $z=0.135$ \citep{Yao2022}; the Tasmanian Devil (AT\,2022tsd) at $z=0.256$, which subsequently displayed extraordinary minute-duration optical flares many weeks after its initial decline, providing direct evidence for a persistent central engine \citep{Ho2023tsd}.  AT\,2023fhn (the Finch) was found at a projected offset in excess of that seen in any previous LFBOT \citep{Chrimes2024}. The most recent well-studied example is AT\,2024wpp \citep[the whippet;][]{2025ApJ...993L...6N,2026ApJ...997L..10L,at2024wppPerley}, at the relatively nearby redshift of $z=0.0868$. This is the most luminous LFBOT known ($L_{\rm bol}\simeq2\times10^{45}$\,\ergs), and has been interpreted as a central engine that produces mildly relativistic winds which interact with the circumburst material \citep{2025ApJ...993L...6N,at2024wppPerley}. 

\cite{Sevilla2026} added analysis of six more LFBOTs to this small, but growing, sample. There is some diversity in host galaxy mass, offset, and multi-wavelength behaviour, while all share the core observational signatures that define the class. \cite{Sevilla2026} highlight the consistent radio behaviour, implying similar circumburst media around each terminal event. Physical scenarios include the direct collapse of a stripped massive star to a compact remnant with a hyper-accreting disc, the merger of a compact object with a stripped companion star, and the tidal disruption of a star by an intermediate-mass black hole \citep{2019ApJ...872...18M,Perley2019,Metzger2022,2025arXiv251209017W}. The persistent UV emission of \cow, 5 years after the event argues for the existence of an accretion disk formed during the tidal disruption scenario \citep{2023MNRAS.525.4042I, 2025MNRAS.544L.108I}.

Here we report the discovery and initial follow-up of \qfm, a luminous fast blue optical transient identified in the ZTF alert stream. In this letter we adopt AB magnitudes throughout, and a flat $\Lambda$CDM cosmology with $H_0=70\,\mathrm{km\,s^{-1}\,Mpc^{-1}}$ and $\Omega_{\rm m}=0.3$.

\section{Discovery}
\label{sec:24qfm-discovery}

\qfm\ (ZTF24aaxhxhf) was first detected in the ZTF alert stream on 24 July 2024 (MJD~60515.4) as part of the public survey, with the most recent non-detection recorded 0.98\,d earlier. The ZTF detections are assimilated into objects by the ZTF team themselves and also distributed (immediately if part of the public survey) to third-party brokers who have their own object assimilation, classification, added-value and discovery announcement processes. The transient's discovery was reported to the Transient Name Server by the ALeRCE broker \citep{2024TNSTR2624....1F} as a young supernova candidate, on the basis of its stamp classifier \citep{Forster2021} and rising lightcurve. Independently, we flagged the source as a rapidly evolving extragalactic transient with the Fastfinder annotator on the Lasair broker \citep{Fulton2026PhDThesis}, and reported it publicly \citep{Fulton2024}.

Lasair is the UK Community Broker for the Legacy Survey of Space and Time (LSST) on the Vera C.\ Rubin Observatory \citep{Williams2024}. Ahead of the start of Rubin Observatory operations, Lasair ingests and serves the public ZTF alert stream, allowing users to construct bespoke filters and to run their own annotator code against the incoming alert stream in near-real time \citep{2019RNAAS...3...26S}. The Fastfinder annotator used here \citep{Fulton2026PhDThesis} is designed specifically to flag transients with rapid photometric evolution -- both fast rises and fast declines -- to prioritise follow-up that is needed on a rapid timescale. Fastfinder is now  being integrated into the Lasair LSST annotation system to flag objects in Rubin's data stream \citep{Williams2024}. \qfm\ was flagged by Fastfinder on the basis of ZTF photometry showing a decline, $\mathrm{d}g / \mathrm{d}t = 0.37\pm0.14$\,mag\,d$^{-1}$, with a comparable decline rate in $r$-band, and a blue colour at the epochs available.

\begin{figure}
    \centering
    \includegraphics[width=0.8\columnwidth]{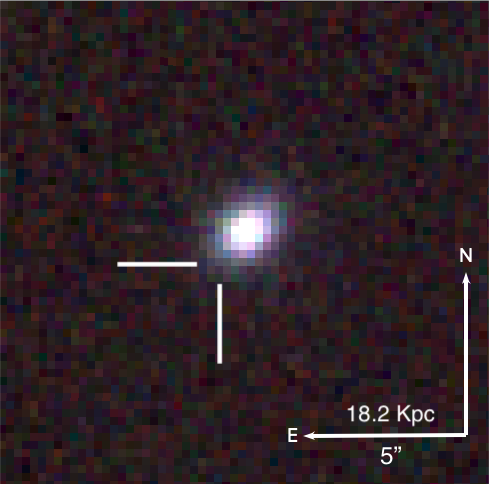}
    \includegraphics[width=0.8\columnwidth]{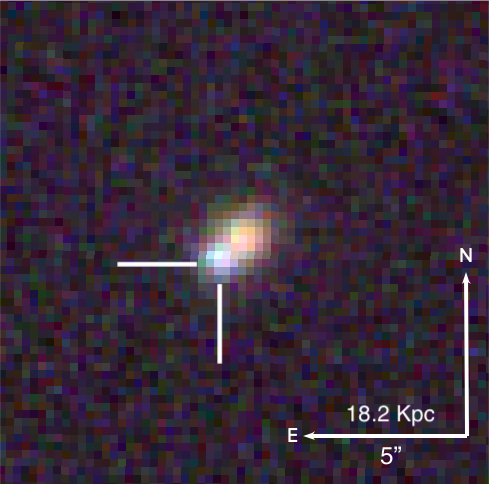}
    \caption{
        Pan-STARRS $riz$-band colour composite images of the host galaxy of \qfm. The location of \qfm\ is indicated by cross hairs.
        \textit{Top:} host galaxy image prior to the transient's appearance.
        \textit{Bottom:} same as {\it Top}, but after the transient's appearance (MJD~60522; 4\,d post peak).
    }
    \label{fig:host_obj_colour_images}
\end{figure}

\qfm\ is located $0.73\arcsec$ south and $1.17\arcsec$ east of an extended source catalogued in the Sloan Digital Sky Survey (SDSS) and Pan-STARRS as SDSS~J232123.40+115632.7 (shown in Figure~\ref{fig:host_obj_colour_images}), for which the SDSS Data Release 15 gives a photometric redshift, $z_{\rm phot}=0.19\pm0.05$. The SDSS photometric redshift and the transient's peak brightness, $m_g=19.51$\,mag, implied a peak absolute magnitude of $M\approx-20.3$~mag, already placing \qfm\ in  the region occupied by \cow-like events. The full lightcurve and access to the Fastfinder annotator output for this object remain publicly available on the Lasair broker pages for ZTF24aaxhxhf.\footnote{\url{https://lasair-ztf.lsst.ac.uk/objects/ZTF24aaxhxhf}} Follow-up was triggered immediately following the public announcement, and photometric confirmation of the rapid decline was obtained and reported in subsequent AstroNotes \citep[e.g.][]{Gillanders2024a}.

\section{Data}
\label{sec:24qfm-data}

\subsection{UV and Optical Photometry}
\label{sec:24qfm-phot}

We obtained a series of imaging observations  with the Pan-STARRS1 and Pan-STARRS2 telescopes \citep{Chambers2016} on Haleakal\=a,  sampling the declining phase of the lightcurve from approximately $7 - 18$\,d after the first ZTF detection, in the \textit{griz} filters, using the standard Pan-STARRS image-differencing pipeline \citep{magnier2020a} to remove the contribution of the underlying host galaxy. The Pan-STARRS photometry confirmed the rapid, blue decline showing \qfm to fade by several tenths of a magnitude per day, and remaining bluer than $g-r\approx-0.3$~mag throughout the interval covered. No evidence for a secondary, radioactive-powered peak is seen in any band over this baseline, consistent with the behaviour established for \cow\ and its analogues \citep{2018ApJ...865L...3P,Perley2021}.

Seven epochs of LOT observations were conducted and reduced using a custom-built pipeline following standard procedures, before being template subtracted and having PSF photometry measured using the python tool \textsc{AutoPhOT} \citep{2022A&A...667A..62B}. Similarly, eight epochs of LT observations were conducted and reduced using the internal IO:O pipeline, before being template subtracted and having a PSF forced onto the resulting difference images using \textsc{photometry-sans-frustration} \citep{2023ApJ...954L..28N}. Count measurements from the LOT and LT were  calibrated against Pan-STARRS1 $3\pi$ survey field stars.

At the position of \qfm, we recovered ten epochs of ZTF $gr$-band forced photometry \citep{2023arXiv230516279M}, including two non-detections prior to discovery, and three epochs of ATLAS $o$-band forced photometry \citep{Tonry2018_ATLAS,Smith2020_ATLAS}, obtained at similar times to the ZTF $r$-band peak. In ATLAS, the individual 30\,s nightly exposures were stacked using an inverse-variance weighted mean to improve the signal-to-noise ratio of each epoch.

We downloaded all available \textit{Swift} UVOT data for \qfm\ to date from the HEASARC Data Archive, obtaining ten epochs approximately $8 - 13$\,d after the first ZTF detection in the $UVW2$ and $UVM2$ filters. No template subtraction was applied to the \textit{Swift} data, and so host galaxy flux contamination is a potential concern -- particularly where the measured transient flux is comparable to, or fainter than, the limits we can place on the host. Upon visual inspection of the UVOT images, a very faint host  is visible at the position of \qfm. 

\begin{figure}
    \centering
    \includegraphics[width=0.8\columnwidth]{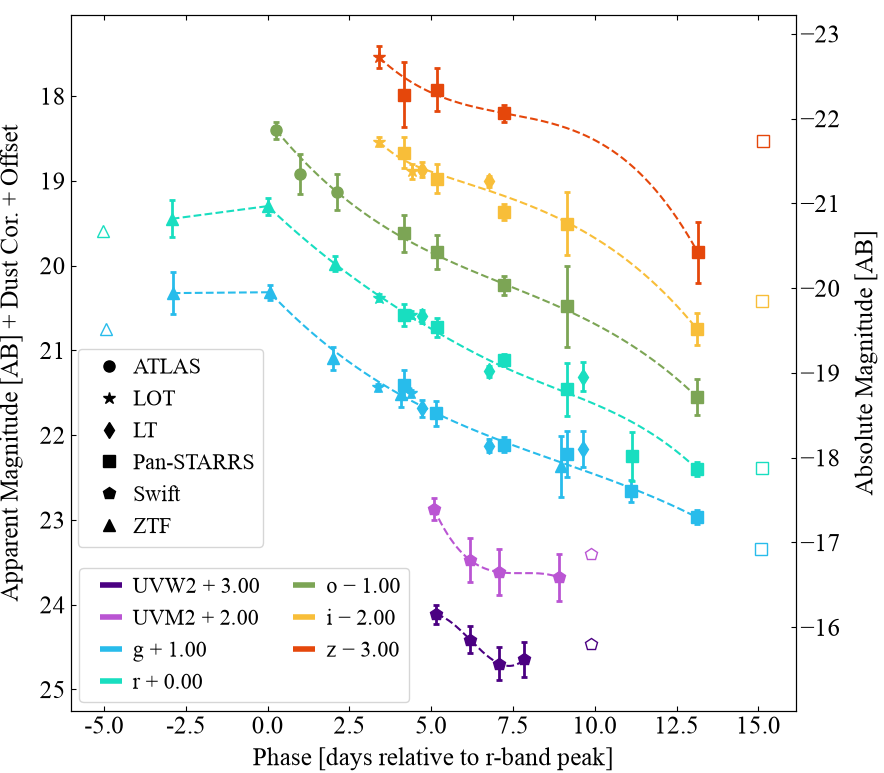}
    \caption{
        Multi-filter lightcurve of \qfm. Filled markers denote detections whereas open markers indicate $3 \sigma$ non-detections. All magnitudes have been corrected only for Galactic dust extinction. Dashed lines show a smoothed spline interpolation to each filter's lightcurve. Phase is given relative to the epoch of $r$-band peak brightness measured by ZTF.
    }
    \label{fig:24qfm-lightcurve}
\end{figure}

Figure~\ref{fig:24qfm-lightcurve} shows the resulting multi-instrument, multi-filter lightcurve of \qfm, combining photometry from ATLAS, LOT, LT, Pan-STARRS, \textit{Swift} and ZTF across seven filters spanning UV to optical wavelengths. Combining our LOT, LT, and Pan-STARRS observations with the ATLAS and ZTF forced photometry, we constrain the rise time above half-maximum to a few days, and the decline rate to $\mathrm{d}g/\mathrm{d}t\approx0.3$\,mag\,d$^{-1}$, essentially unchanged from the discovery epoch estimate and the values reported in follow-up AstroNotes \citep{Fulton2024, Gillanders2024a}.

\subsection{Spectroscopy and redshift estimation} 
\label{sec:24qfm-GMOS}

\begin{figure*}
    \centering
    \includegraphics[width=0.8\linewidth]{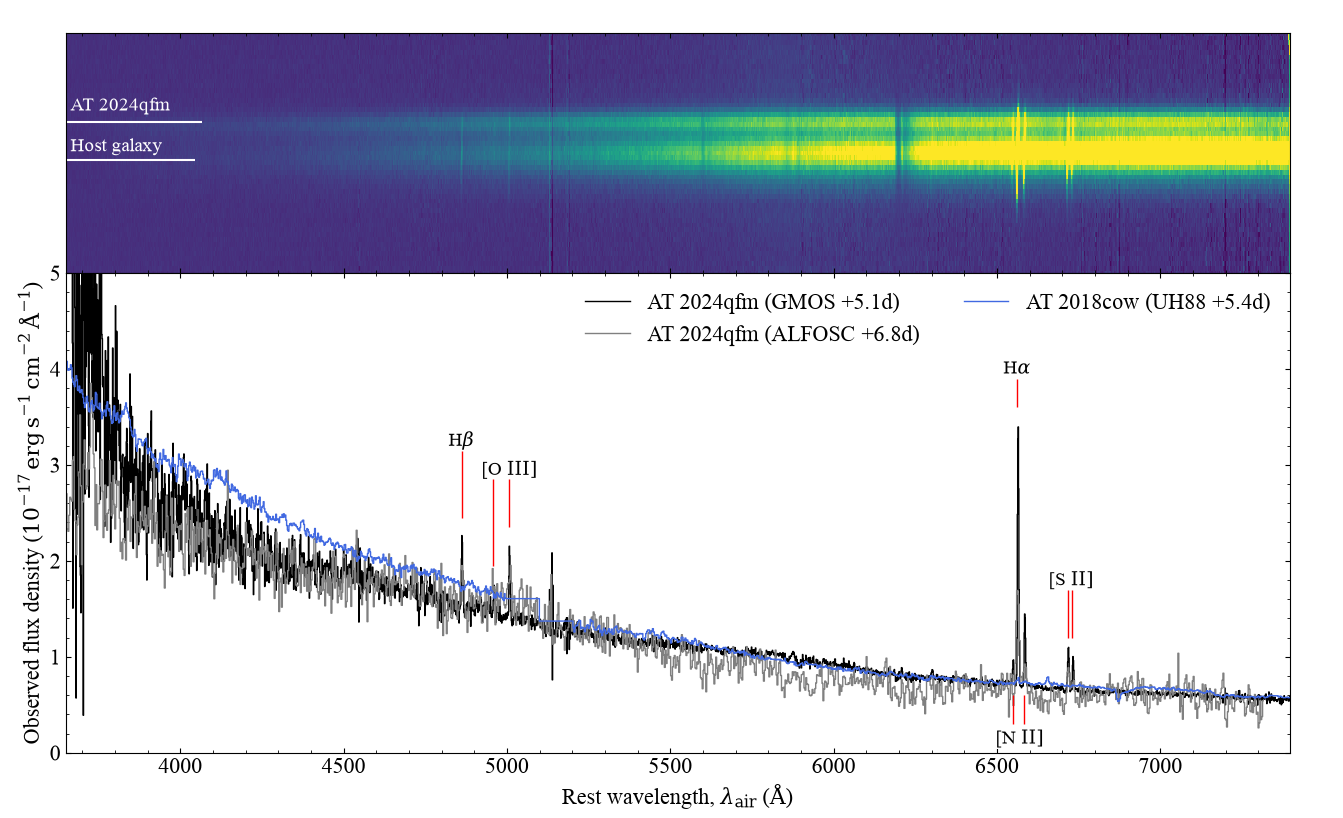}
    \caption{
        \textit{Upper panel:} 2D Gemini/GMOS spectrum of \qfm\ and its host galaxy. The location of the host galaxy and transient traces are marked.
        \textit{Lower panel:} 1D spectra of \qfm\ at two epochs: Gemini/GMOS at +5.1\,d and NOT/ALFOSC at +6.8\,d from $r$-band peak, compared against \cow\ (flux-scaled to match AT 2024qfm's median flux density over $5200 - 6200$\,\AA). All spectra have been corrected for Milky Way extinction and are shown in the host galaxy's rest frame (z = 0.2267). Prominent emission features from which we extract a redshift estimate are marked and labelled.
    }
    \label{fig:24qfm_Gemini_spectrum}
\end{figure*}

We obtained long-slit spectroscopy of \qfm\ with the Gemini Multi-Object Spectrograph (GMOS) on Gemini-North, under the target-of-opportunity programme GN-2024A-Q-128 (PI:~Huber). Observations began on 1 August 2024 at 10:57:58~UT, $\simeq 8$~days after the first significant ZTF detection ($\simeq 5$~days after the reported $r$-band peak), and comprised of $4 \times 1200$\,s exposures with the R400 grating sampling the observed wavelength range $\approx 4200 - 9100$\,\AA\ \citep[as first reported by][]{Gillanders2024a}. The images were nodded along the slit to spatially remove the sky lines, and shifted in wavelength to cover the GMOS chip gaps. We also obtained spectra of \qfm\ using the Alhambra Faint Object Spectrograph and Camera (ALFOSC) on the 2.56-m Nordic Optical Telescope (NOT) on 03 August 2024 using a 1.0\arcsec\ slit with grism 4. The spectrum was obtained under good observing conditions. We perform cosmic ray rejection on the raw data using {\tt AstroSCRAPPY} before using standard routines within {\tt IRAF} and {\tt PYRAF} to extract, wavelength calibrate and flux calibrate the spectrum.

Figure~\ref{fig:24qfm_Gemini_spectrum} presents the 2D and 1D GMOS spectra of \qfm, flux-calibrated against contemporaneous Pan-STARRS photometry, corrected for Milky Way foreground extinction, and de-redshifted to the rest frame of the host galaxy (see below). Alongside this, we plot a spectrum obtained with ALFOSC on the Nordic Optical Telescope $\simeq 1.7$\,d later, together with a spectrum of \cow\ \citep{2018ApJ...865L...3P} obtained at a comparable phase relative to its own $r$-band peak. In all three cases the spectra are characterised by a blue, largely featureless continuum, with the only identifiable features being narrow host-galaxy emission lines. We find no evidence for broad, transient-intrinsic spectral features at any of these epochs.

Extracting the spectrum of the host galaxy reveals a set of narrow nebular emission lines, including H$\alpha$, H$\beta$, [O\,{\sc iii}]~$\lambda \lambda 5006.8$, 4958.9, [N\,{\sc ii}]~$\lambda \lambda 6548.1$, 6583.5 and [S\,{\sc ii}]~$\lambda \lambda 6716.4$, 6730.8, all consistent with a common redshift, $z = 0.2267 \pm 0.0002$, which is within the error of the SDSS DR15 photometric redshift, and notably very compatible with the Legacy Survey DR9 photometric redshift of $z=0.234\pm0.049$.  The transient \qfm\ is spatially offset by 1.1$''$ from the host and the nebular emission lines, which indicate ongoing star-formation activity, clearly extend to the transient's position at a projected radius of 4.0\,kpc, and significantly beyond the host's detectable continuum emission. At the position of \qfm, the emission lines show a velocity offset, $v = +99$\,\kms. This is compatible with the rotational velocity expected for an $M_{\rm baryonic}\simeq10^{10}$\,\msol\ galaxy measure at a physical radius of between 4-5\,kpc (see Section~\ref{sec:24qfm-host}). 

\section{Host Galaxy analysis}
\label{sec:24qfm-host}

\begin{table} 
    \centering
    \caption{
        Physical properties of the host galaxy of \qfm\ derived from \texttt{Bagpipes} SED fitting. Values are the medians of the marginalised posteriors, with uncertainties giving the 16th--84th percentile credible interval.
    }
    \label{tab:24qfm_host_properties}
    \begin{tabular}{lc}
        \hline
        Property & Value \\
        \hline
        $\log_{10}(M_\star/{\rm M}_\odot)$ & $10.22^{+0.04}_{-0.05}$ \\[2pt]
        SFR $[{\rm M}_\odot\,\mathrm{yr^{-1}}]$ & $1.43^{+0.45}_{-0.31}$ \\[2pt]
        $\log_{10}(\mathrm{sSFR}\,[\mathrm{yr^{-1}}])$ & $-10.06^{+0.13}_{-0.12}$ \\[2pt]
        $A_V$ [mag] & $0.16^{+0.06}_{-0.05}$ \\[2pt]
        Mass-weighted age [Gyr] & $7.07^{+0.84}_{-0.84}$ \\
        \hline
    \end{tabular}
\end{table}

We retrieve image cutouts from GALEX ($FUV$, $NUV$), SDSS ($u$), PanSTARRS ($grizy$), UKIDSS ($JHK$), and unWISE ($W1$--$W4$) to analyse the the host galaxy of AT\,2024qfm  using the \texttt{HostPhot} package \citep{HostPhot}. A galaxy Kron aperture was measured on the co-added Pan-STARRS $riz$ images. This reference aperture was rescaled by a per-band factor that maximises the signal-to-noise ratio of the enclosed flux; this results in significantly larger apertures for the coarse-resolution bands (e.g., the $W1$ aperture is $\sim2.63\times$ larger than the Pan-STARRS $r$-band aperture). We further add a photometric error floor of 0.05\,mag in quadrature to all bands to account for calibration systematics, and Galactic foreground extinction is corrected using the dust maps of \citet{2011ApJ...737..103S} and the extinction law of \citet{Fitzpatrick1999} with $R_V=3.1$.

We model the resulting spectral energy distribution with the Bayesian SED-fitting code \texttt{Bagpipes} \citep{Bagpipes}, with a non-parametric, continuity star-formation history \citep{Leja2019}, in which the SFH is described by the star-formation rate in seven lookback-time bins. The first two bins are fixed to $0$--$30$\,Myr and $30$--$100$\,Myr, the most recent bin extends to $90\%$ of the age of the Universe at the host redshift, and the remaining bins are spaced logarithmically in between; the final bin spans up to the age of the Universe at $z=0.2267$. A Student's-$t$ prior \citep[scale $=0.3$, $\nu=2$;][]{Leja2019} is placed on the logarithmic ratio of the SFR between adjacent bins, which disfavours sharp, unphysical transitions in the SFH. The priors over which the Bayesian sampling is performed are described in Appendix\,\ref{app:host} and the posterior distribution is sampled using the \texttt{MultiNest} nested-sampling algorithm \citep{Feroz2009} via its \texttt{PyMultiNest} interface \citep{Buchner2016}.  We report the median and $16$th--$84$th percentile credible intervals of the marginalised posteriors for derived galaxy quantities in Table \ref{tab:24qfm_host_properties} (with a corner plot provided in Appendix\,\ref{app:host}). 

The H$\alpha$ and H$\beta$ emission line fluxes from the host (after subtraction of the continuum) are $8.7$ and $0.39\times10^{-16}$\,erg\,s$^{-1}$\,cm$^{-2}$ respectively. This Balmer decrement implies an extinction at 6562\AA\ rest-frame of $A_{\rm H \alpha}=1.2$, an H$\alpha$ luminosity of $L_{\rm H\alpha}=4.0\times10^{41}$\,erg\,s$^{-1}$ and a star-formation rate of 2.1\msol\,yr$^{-1}$ from \cite{KE2021}, in reasonable agreement with the \texttt{Bagpipes} SED fit and lower than than that from the dustier SED fit of  \citet{Sevilla2026}. The host  stellar mass and SFR at the higher end of the distribution of these properties amongst LFBOT hosts but is not an outlier \citep{at2024wppPerley,Sevilla2026}.  

\section{Discussion and comparison with other sources}
\label{sec:24qfm-discussion}

\begin{figure}
    \centering
    \includegraphics[width=\columnwidth]{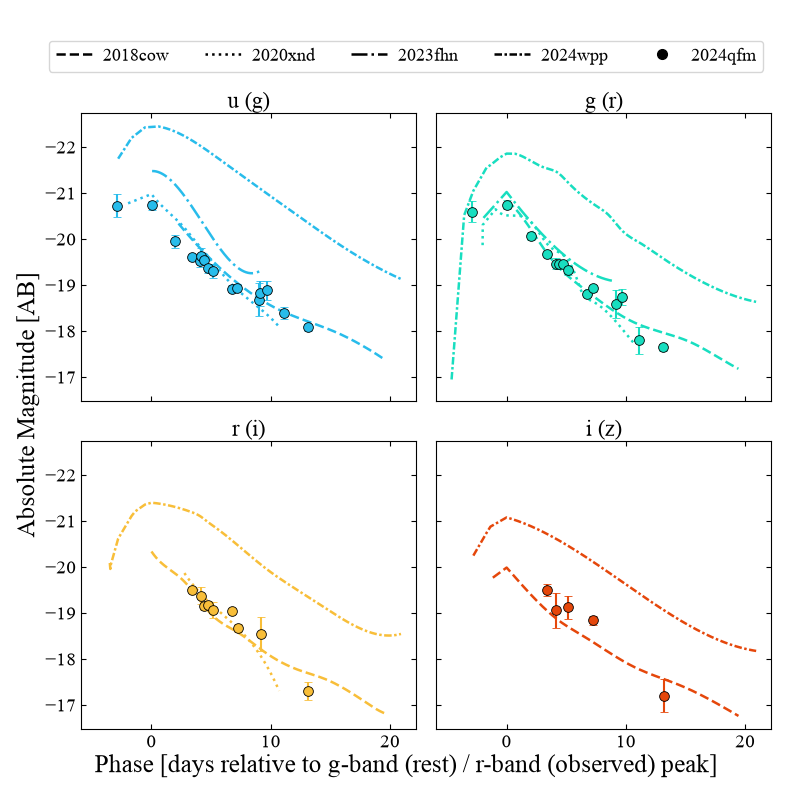}
    \caption{
        Absolute-magnitude lightcurve of \qfm\ compared to four known LFBOTs. Phase is relative to peak in the observed $r$-band, equivalently the rest-frame $g$-band at \qfm's redshift (panel titles give the rest-frame filter with its observed-frame counterpart in parenthesis). 
    }
    \label{fig:24qfm-lightcurveComp}
\end{figure}

\qfm\ shares the defining observational characteristics of the LFBOT class: a rise and decline confined to a few days, a blue and largely featureless spectrum near peak, and a peak luminosity, $M_g\simeq-21.0$. Figure~\ref{fig:24qfm-lightcurveComp} compares the absolute-magnitude evolution of \qfm\ to that of known LFBOTs: \cow\ \citep{2018ApJ...865L...3P, Perley2019}, AT\,2020xnd \citep{Perley2021}, AT\,2023fhn \citep{Sevilla2026}, and AT\,2024wpp \citep{at2024wppPerley} across four rest-frame filters. The absolute magnitudes for \cow\ and AT\,2024wpp were plotted without any colour or $K-$correction, apart from a foreground extinction term. However, for AT\,2020xnd, AT\,2023fhn, and \qfm, we convert their observer-frame magnitudes to respective rest-frame absolute magnitudes after correcting for foreground dust extinction, distance and an approximate $K-$correction \citep{2002astro.ph.10394H}. As these all possess $z\simeq0.23 - 0.24$, a correction of  $K=-2.5\log(1+z)$ is added when calculating the absolute magnitude, which shifts observer-frame $g$-band AB magnitudes to rest-frame $u$-band. The redder filters are all similarly shifted to bluer rest-frame filters in the $ugriz$ PS/SDSS-like sequence. 

The rest-frame $ugri$ absolute magnitudes and colours of \qfm\ are remarkably similar to both \cow\ and AT\,2020xnd. By implication, their effective temperatures and the size and velocity of their emitting region's radii are also very similar. This is further demonstrated in Figure~\ref{fig:24qfm_Gemini_spectrum}, where the spectra of \qfm\ show an identical featureless blue continuum to \cow. As such, the UV+optical data of \qfm\ does not offer significantly new insights into the nature of LFBOTs, as model fits would recover similar conclusions to those already proposed to explain the \cow\ data sets. The optical data alone cannot distinguish between circumstellar shock models \citep{2024ApJ...972..140K} and accretion powering in a central engine \citep{2019MNRAS.485L..83Q}, or, potentially, the merger of black holes and Wolf-Rayet stars \citep{Metzger2022}. Further X-ray and radio analysis will be presented on \qfm\ (see \citealt{Sevilla2026} for details). 

\qfm\ does offer another object to add to a small (but growing) sample of LFBOTs, and its discovery method points to how more can be found. \cite{Sevilla2026} highlighted the noticeable projected offset from its host galaxy, which we also confirm, and is evident in Figure~\ref{fig:host_obj_colour_images}. This is intermediate between the small ($\lesssim2$\,kpc) offsets typical of CSS161010, \cow, AT\,2018lug, AT\,2020mrf and AT\,2020xnd, and the much larger offsets recently reported for AT\,2022tsd and, especially, AT\,2023fhn.  The emerging picture is that LFBOT host-galaxy offsets span a wider range than was apparent from the first few examples of the class. In comparison to the LFBOT host population of \citet{Sevilla2026}, our analysis finds that the host of \qfm\ lies toward the massive end, while its actively star-forming nature is typical of the sample. It also hosts the oldest stellar population of known LFBOT hosts, with a dust content toward the lower end of the distribution; together, these observations are indicative of a more evolved stellar population than in typical LFBOT hosts.

\section{Conclusion and Outlook}
\label{sec:24qfm-conclusion}

\qfm\ was independently flagged as a rapidly evolving extragalactic transient by our Fastfinder annotator running on the ZTF alert stream through Lasair, on the basis of a fast decline ($\mathrm{d}g/\mathrm{d}t = 0.37\pm0.14$\,mag\,d$^{-1}$) and blue colour in the early forced-photometry epochs, and was reported publicly ahead of spectroscopic classification. Gemini/GMOS and NOT/ALFOSC spectroscopy revealed narrow host-galaxy nebular emission lines at a common redshift of $z=0.2267\pm0.0002$, with the transient itself showing only a blue, featureless continuum closely matching a contemporaneous-phase spectrum of \cow. At this redshift, \qfm\ reached a peak rest-frame magnitude of $M_g=-21.0$. The rapid rise and decline, the persistently blue and featureless spectrum, the peak luminosity, and the rest-frame $ugri$ colours and inferred temperature closely tracking both \cow\ and AT\,2020xnd, place \qfm\ firmly within the LFBOT class, adding another member to the still-small sample of confirmed events. The data gathered show it to be almost identical to \cow. 

This discovery demonstrates both the potential and limitations of finding fast transients in LSST survey data with the Lasair broker. The Fastfinder algorithm, also flagged another clear LFBOT candidate AT\,2024kth,\footnote{\url{https://lasair-ztf.lsst.ac.uk/objects/ZTF24aarbequ}} in a diffuse host at a photometric redshift of $z=0.16 \pm 0.06$. The fast decline ($dm_{g}/dt = -0.26\pm0.04$\,mag\,day$^{-1}$) and bright absolute magnitude ($M_g = -20.9 \pm 0.9$) indicate similarity with the LFBOT population, but no spectroscopic confirmation was taken at the time \citep{Fulton2026PhDThesis}. Taken together, these two discoveries would push the estimated rates in \citep{at2024wppPerley}, based on only 3 LFBOTS in ZTF, up by 66\%. 

Despite the efficiency of the Fastfinder annotator, there is a practical latency involved in confirming  LFBOT candidates. The transient was initially detected and reported by ZTF on 27 July 2024, yet it took a further five days before we were able to confirm it spectroscopically as an LFBOT event on 1 August 2024  \citep{Fulton2024,Gillanders2024a}. This delay was due to the requirement of a second epoch to measure the fast decline, the subsequent Slack alerting of the Fastfinder annotation \citep{Fulton2026PhDThesis}, manual review required before public announcement, and then latency in the Gemini trigger. LSST offers the opportunity to discover many LFBOTs from candidate selection based on the combination of rapid decline and colour, but we highlight that recognising them early enough to trigger multi-wavelength campaigns will be challenging. The LSST Wide--Fast--Deep (WFD) simulated survey \citep{2022ApJS..258....1B} has $5 \sigma$ single-visit limiting magnitudes, $m_r \leq 24$, $m_i \leq 23.4$. To measure the observed decline of  $\sim0.4$\,mag\,d$^{-1}$ in rest-frame $u$- or $g$-band, over the typical $\sim 5$~day inter-night cadence in the same band \citep{2022ApJS..258....1B}, requires an observed peak of $m_r \leq 22$ and $m_i \leq 21.4$. Adopting a rest-frame characteristic peak absolute magnitude of $M_g =-21$ (or similar for rest-frame $u$; see Figure~\ref{fig:24qfm-lightcurveComp}), sets a redshift limit to detect the source at peak {\em and} measure the decline with some significance. Again, allowing for a simple $K-$correction of $2.5 \log(1+z)$, an observed peak of $m_r\leq22$ implies a distance limit of $\mu \approx 42.7$, or a maximum redshift of $z\approx0.8$ (for $m_i \leq 21.4$ the limit is $z\approx0.65$). The LSST $r$ and $i$ filters would approximately correspond to rest-frame $u$- and $g$-bands \citep[effective wavelengths from][]{2022ApJS..258....1B}, meaning LSST could comfortably identify an  LFBOT via decline rate and $g-r$ colour alone, within these limits. This opens up a co-moving volume $\sim 10 \times$ larger than that probed by ZTF for LFBOTs ($z\lesssim0.3$), potentially meaning we could detect 1 per month, rather than the current rate of just over 1 per year. However, the challenge remains in identifying them early enough as LFBOTs for multi-facility follow-up; this will effectively be set by LSST's chosen cadence. An interesting possibility for LSST is identifying LFBOTS fading in the intra-night third exposure that is planned for a portion of the sky  \citep{2019PASP..131f8002B}. A 6\,hr time difference would potentially highlight a 0.1\,mag fade, which may be detectable. 

\bibliographystyle{mnras}
\bibliography{at2024qfm}

\section*{Data Availability}

The UV and optical photometric data are provided in Appendix~\ref{app:2024qfm_photometrytable}. The spectral data are publicly available on the Weizmann Interactive Supernova Data Repository (WISeREP).\footnote{\url{https://www.wiserep.org}} All of these data are being made public at the time of submission. The ATLAS and ZTF data are available through the ATLAS forced photometry server\footnote{\url{https://fallingstar-data.com}} and the Lasair ZTF broker,\footnote{\url{https://lasair-ztf.lsst.ac.uk}} respectively. \textsc{photometry-sans-frustration}\footnote{\url{https://github.com/mnicholl/photometry-sans-frustration}} is a  python tool utilising the \textit{Astropy} and \textit{Photutils} packages \citep{2023ApJ...954L..28N}. The LOT custom-built photometry pipeline is publicly available.\footnote{\url{https://hdl.handle.net/11296/98q6x4}} The NOT data reduction made use of {\tt AstroSCRAPPY},\footnote{\url{https://zenodo.org/records/1482019}} and standard routines within {\tt IRAF}\footnote{\url{https://iraf-community.github.io}} and {\tt PYRAF}.\footnote{\url{https://github.com/iraf-community/pyraf}} 

\section*{Acknowledgements}

Lasair is supported by the UKRI Science and Technology Facilities Council and is a collaboration between the University of Edinburgh (grant ST/N002512/1) and Queen's University Belfast and University of Oxford (grant ST/N002520/1) within the \href{https://www.lsst.ac.uk/}{LSST:UK} Science Consortium. 

ZTF is supported by National Science Foundation grant AST-1440341 and a collaboration including Caltech, IPAC, the Weizmann Institute for Science, the Oskar Klein Center at Stockholm University, the University of Maryland, the University of Washington, Deutsches Elektronen-Synchrotron and Humboldt University, Los Alamos National Laboratories, the TANGO Consortium of Taiwan, the University of Wisconsin at Milwaukee, and Lawrence Berkeley National Laboratories. Operations are conducted by COO, IPAC, and UW.

Pan-STARRS is primarily funded to search for near-Earth asteroids through NASA grants NNX08AR22G and NNX14AM74G. The Pan-STARRS science products were made possible through the contributions of the University of Hawai'i Institute for Astronomy, the Queen's University Belfast and the University of Oxford. 

ATLAS is primarily funded to search for near earth asteroids through NASA grants NN12AR55G, 80NSSC18K0284, and 80NSSC18K1575 (under the guidance of Lindley Johnson and Kelly Fast); by-products of the NEO search include images and catalogues from the survey area. The ATLAS science products have been made possible through the contributions of the University of Hawaii Institute for Astronomy, the Queen’s University Belfast, the Space Telescope Science Institute, and the South African Astronomical Observatory.

The data presented here were obtained in part with ALFOSC, which is provided by the Instituto de Astrofisica de Andalucia (IAA) under a joint agreement with the University of Copenhagen and NOT, under programme P69-014 (PI: Angus).

S. J. Smartt, S. Srivastav, and K. W. Smith acknowledge funding from STFC Grants ST/Y001605/1, ST/X001253/1, a Royal Society Research Professorship and the Hintze Family Charitable Foundation. T.-W. Chen and C.-H. Lai acknowledge funding from the Yushan Fellow Program by the Ministry of Education, Taiwan (MOE-111-YSFMS-0008-001-P1) and the National Science and Technology Council, Taiwan (NSTC grant 114-2112-M-008-021-MY3). We also acknowledge Amar Aryan for processing the LOT data. MP acknowledges support from a UK Research and Innovation Fellowship (UKRI1062). Parts of this research were supported by the Australian Research Council Centre of Excellence for Gravitational Wave Discovery (OzGrav), through project number CE230100016. S. Yang acknowledges funding from the National Natural Science Foundation of China under grant No. 12303046, the Startup Research Fund of Henan Academy of Sciences No. 242041217, and the Joint Fund of Henan Province Science and Technology R\&D Program No. 235200810057.

\appendix

\section{Photometry data tables}
\label{app:2024qfm_photometrytable}

In Tables~\ref{tab:2024qfm_photometry_optical} and \ref{tab:2024qfm_photometry_uv}, we provide the full photometric dataset for \qfm\ behind Figures~\ref{fig:24qfm-lightcurve} and \ref{fig:24qfm-lightcurveComp}. Table~\ref{tab:2024qfm_photometry_optical} contains the optical photometry obtained from various ground-based telescopes, while Table~\ref{tab:2024qfm_photometry_uv} contains the UV photometric dataset obtained with \textit{Swift}. All measurements pertain to the LFBOT and are given in AB magnitudes and micro-Janskys, uncorrected for Galactic or host dust extinction, with magnitude limits quoted to $3\sigma$. The ground-based measurements have additionally been corrected for host-galaxy contribution, whereas the \textit{Swift} data have not. The total exposure time reported for each night is typically the sum of the individual sub-exposures combined. Where the exposure time for an $o$-filter measurement is marked with an asterisk (*), the measurement instead represents a composite of the $r$- and $i$-bands, as described by \cite{Tonry2018_ATLAS}. The phase quoted is with respect to the peak brightness epoch measured in the ZTF $r$-band.

\onecolumn
\begin{longtable}{ccccccccc}
    \caption{Optical photometry of \qfm.}
    \label{tab:2024qfm_photometry_optical} \\
    
    \hline\hline
    MJD & Phase (d) & Filter & Flux ($\mu$Jy) & dFlux ($\mu$Jy) & Mag (AB) & dMag (AB) & Exptime (s) & Instrument \\
    \hline
    \endfirsthead
    
    \multicolumn{9}{c}{{\tablename\ \thetable{} -- continued from previous page}} \\
    \hline\hline
    MJD & Phase (d) & Filter & Flux ($\mu$Jy) & dFlux ($\mu$Jy) & Mag (AB) & dMag (AB) & Exptime (s) & Instrument \\
    \hline
    \endhead
    
    \hline
    \multicolumn{9}{r}{{Continued on next page}} \\
    \endfoot
    
    \hline\hline
    \endlastfoot
    
    60513.324 & $-5.022$            & $r$ & 4.1  & 9.2  & $>19.7$ & --   & 30   & ZTF        \\
    60513.399 & $-4.947$            & $g$ & 9.9  & 8.3  & $>19.9$ & --   & 30   & ZTF        \\
    60515.411 & $-2.935$            & $r$ & 54.6 & 11.8 & 19.56   & 0.22 & 30   & ZTF        \\
    60515.442 & $-2.904$            & $g$ & 58.9 & 14.9 & 19.48   & 0.25 & 30   & ZTF        \\
    60518.346 & $\phantom{-}0.000$  & $r$ & 62.7 & 6.0  & 19.41   & 0.10 & 30   & ZTF        \\
    60518.401 & $\phantom{-}0.055$  & $g$ & 59.2 & 5.2  & 19.47   & 0.09 & 30   & ZTF        \\
    60518.601 & $\phantom{-}0.255$  & $o$ & 57.8 & 6.0  & 19.50   & 0.10 & 120  & ATLAS      \\
    60519.346 & $\phantom{-}1.000$  & $o$ & 35.7 & 8.4  & 20.02   & 0.24 & 120  & ATLAS      \\
    60520.350 & $\phantom{-}2.004$  & $g$ & 28.7 & 4.1  & 20.25   & 0.14 & 30   & ZTF        \\
    60520.390 & $\phantom{-}2.044$  & $r$ & 33.3 & 3.1  & 20.09   & 0.09 & 30   & ZTF        \\
    60520.472 & $\phantom{-}2.126$  & $o$ & 29.5 & 6.1  & 20.23   & 0.21 & 120  & ATLAS      \\
    60521.730 & $\phantom{-}3.384$  & $g$ & 21.2 & 0.8  & 20.59   & 0.04 & 900  & LOT        \\
    60521.734 & $\phantom{-}3.388$  & $r$ & 23.2 & 0.8  & 20.49   & 0.04 & 900  & LOT        \\
    60521.747 & $\phantom{-}3.401$  & $i$ & 20.5 & 1.0  & 20.62   & 0.05 & 900  & LOT        \\
    60521.750 & $\phantom{-}3.404$  & $z$ & 20.6 & 2.7  & 20.61   & 0.13 & 900  & LOT        \\
    60522.423 & $\phantom{-}4.077$  & $g$ & 19.5 & 3.0  & 20.67   & 0.15 & 30   & ZTF        \\
    60522.504 & $\phantom{-}4.158$  & $g$ & 21.7 & 3.7  & 20.56   & 0.17 & 200  & Pan-STARRS \\
    60522.507 & $\phantom{-}4.161$  & $r$ & 19.2 & 2.5  & 20.69   & 0.13 & 200  & Pan-STARRS \\
    60522.508 & $\phantom{-}4.162$  & $o$ & 18.7 & 4.1  & 20.72   & 0.22 & 200* & Pan-STARRS \\
    60522.509 & $\phantom{-}4.163$  & $i$ & 18.2 & 3.3  & 20.75   & 0.18 & 200  & Pan-STARRS \\
    60522.512 & $\phantom{-}4.166$  & $z$ & 13.8 & 5.2  & 21.05   & 0.38 & 200  & Pan-STARRS \\
    60522.733 & $\phantom{-}4.387$  & $g$ & 19.8 & 0.7  & 20.66   & 0.04 & 900  & LOT        \\
    60522.737 & $\phantom{-}4.391$  & $r$ & 19.3 & 0.7  & 20.69   & 0.04 & 900  & LOT        \\
    60522.756 & $\phantom{-}4.410$  & $i$ & 14.8 & 1.3  & 20.97   & 0.09 & 900  & LOT        \\
    60523.062 & $\phantom{-}4.716$  & $g$ & 16.7 & 1.6  & 20.84   & 0.10 & 270  & LT         \\
    60523.066 & $\phantom{-}4.720$  & $r$ & 18.8 & 1.5  & 20.71   & 0.08 & 270  & LT         \\
    60523.070 & $\phantom{-}4.724$  & $i$ & 15.1 & 1.4  & 20.95   & 0.09 & 270  & LT         \\
    60523.506 & $\phantom{-}5.169$  & $g$ & 15.8 & 2.4  & 20.90   & 0.15 & 300  & Pan-STARRS \\
    60523.509 & $\phantom{-}5.163$  & $r$ & 16.7 & 1.8  & 20.84   & 0.11 & 300  & Pan-STARRS \\
    60523.511 & $\phantom{-}5.165$  & $o$ & 15.3 & 3.1  & 20.94   & 0.20 & 300* & Pan-STARRS \\
    60523.513 & $\phantom{-}5.167$  & $i$ & 13.7 & 2.3  & 21.06   & 0.17 & 300  & Pan-STARRS \\
    60523.517 & $\phantom{-}5.171$  & $z$ & 14.6 & 3.7  & 20.99   & 0.25 & 300  & Pan-STARRS \\
    60525.098 & $\phantom{-}6.752$  & $g$ & 11.2 & 0.9  & 21.28   & 0.08 & 450  & LT         \\
    60525.104 & $\phantom{-}6.758$  & $r$ & 10.4 & 0.6  & 21.36   & 0.06 & 450  & LT         \\
    60525.110 & $\phantom{-}6.764$  & $i$ & 13.3 & 0.7  & 21.09   & 0.06 & 450  & LT         \\
    60525.557 & $\phantom{-}7.211$  & $g$ & 11.3 & 1.0  & 21.27   & 0.09 & 600  & Pan-STARRS \\
    60525.564 & $\phantom{-}7.218$  & $r$ & 11.7 & 0.8  & 21.23   & 0.07 & 600  & Pan-STARRS \\
    60525.568 & $\phantom{-}7.222$  & $o$ & 10.7 & 1.2  & 21.33   & 0.11 & 600* & Pan-STARRS \\
    60525.571 & $\phantom{-}7.225$  & $i$ & 9.5  & 0.9  & 21.45   & 0.09 & 600  & Pan-STARRS \\
    60525.578 & $\phantom{-}7.232$  & $z$ & 11.3 & 1.1  & 21.27   & 0.10 & 600  & Pan-STARRS \\
    60527.318 & $\phantom{-}8.972$  & $g$ & 9.0  & 3.2  & 21.52   & 0.36 & 30   & ZTF        \\
    60527.482 & $\phantom{-}9.136$  & $g$ & 10.2 & 2.8  & 21.38   & 0.27 & 600  & Pan-STARRS \\
    60527.489 & $\phantom{-}9.143$  & $r$ & 8.6  & 2.7  & 21.57   & 0.31 & 600  & Pan-STARRS \\
    60527.493 & $\phantom{-}9.147$  & $o$ & 8.5  & 4.1  & 21.58   & 0.48 & 600* & Pan-STARRS \\
    60527.497 & $\phantom{-}9.151$  & $i$ & 8.4  & 3.1  & 21.59   & 0.37 & 600  & Pan-STARRS \\
    60527.994 & $\phantom{-}9.648$  & $g$ & 10.8 & 2.3  & 21.32   & 0.21 & 450  & LT         \\
    60528.000 & $\phantom{-}9.654$  & $r$ & 9.8  & 1.7  & 21.42   & 0.17 & 450  & LT         \\
    60529.460 & $\phantom{-}11.114$ & $g$ & 6.8  & 0.9  & 21.82   & 0.13 & 900  & Pan-STARRS \\
    60529.471 & $\phantom{-}11.125$ & $r$ & 4.1  & 1.2  & 22.36   & 0.29 & 900  & Pan-STARRS \\
    60531.464 & $\phantom{-}13.118$ & $g$ & 5.2  & 0.4  & 22.12   & 0.08 & 900  & Pan-STARRS \\
    60531.475 & $\phantom{-}13.129$ & $r$ & 3.6  & 0.3  & 22.51   & 0.08 & 900  & Pan-STARRS \\
    60531.481 & $\phantom{-}13.135$ & $o$ & 3.2  & 0.7  & 22.65   & 0.21 & 900* & Pan-STARRS \\
    60531.486 & $\phantom{-}13.140$ & $i$ & 2.7  & 0.5  & 22.83   & 0.19 & 900  & Pan-STARRS \\
    60531.497 & $\phantom{-}13.151$ & $z$ & 2.5  & 0.9  & 22.91   & 0.36 & 900  & Pan-STARRS \\
    60533.435 & $\phantom{-}15.089$ & $g$ & 2.8  & 1.2  & $>22.5$ & --   & 900  & Pan-STARRS \\
    60533.445 & $\phantom{-}15.099$ & $r$ & 0.3  & 1.2  & $>22.5$ & --   & 900  & Pan-STARRS \\
    60533.468 & $\phantom{-}15.122$ & $i$ & 2.5  & 1.2  & $>22.5$ & --   & 900  & Pan-STARRS \\
    60533.478 & $\phantom{-}15.132$ & $z$ & 1.0  & 2.7  & $>21.6$ & --   & 900  & Pan-STARRS \\
    
\end{longtable}
\twocolumn

\onecolumn
\begin{longtable}{ccccccccc}
    \caption{UV photometry of \qfm.}
    \label{tab:2024qfm_photometry_uv} \\
    
    \hline\hline
    MJD & Phase (d) & Filter & Flux ($\mu$Jy) & dFlux ($\mu$Jy) & Mag (AB) & dMag (AB) & Exptime (s) & Instrument \\
    \hline
    \endfirsthead
    
    \multicolumn{9}{c}{{\tablename\ \thetable{} -- continued from previous page}} \\
    \hline\hline
    MJD & Phase (d) & Filter & Flux ($\mu$Jy) & dFlux ($\mu$Jy) & Mag (AB) & dMag (AB) & Exptime (s) & Instrument \\
    \hline
    \endhead
    
    \hline
    \multicolumn{9}{r}{{Continued on next page}} \\
    \endfoot
    
    \hline\hline
    \endlastfoot
    
    60523.426 & $\phantom{-}5.080$ & $UVM2$ & 12.5 & 1.6 & 21.16   & 0.13 & 1789 & {\it Swift} \\
    60523.491 & $\phantom{-}5.145$ & $UVW2$ & 10.5 & 1.2 & 21.35   & 0.11 & 2963 & {\it Swift} \\
    60524.537 & $\phantom{-}6.191$ & $UVW2$ & 7.9  & 1.3 & 21.65   & 0.16 & 1570 & {\it Swift} \\
    60524.542 & $\phantom{-}6.196$ & $UVM2$ & 7.1  & 1.8 & 21.77   & 0.26 & 799  & {\it Swift} \\
    60525.424 & $\phantom{-}7.078$ & $UVW2$ & 6.1  & 1.1 & 21.93   & 0.19 & 1570 & {\it Swift} \\
    60525.429 & $\phantom{-}7.083$ & $UVM2$ & 6.2  & 1.7 & 21.91   & 0.27 & 841  & {\it Swift} \\
    60526.190 & $\phantom{-}7.844$ & $UVW2$ & 6.4  & 1.3 & 21.88   & 0.21 & 1084 & {\it Swift} \\
    60527.246 & $\phantom{-}8.900$ & $UVM2$ & 5.9  & 1.7 & 21.97   & 0.28 & 887  & {\it Swift} \\
    60528.217 & $\phantom{-}9.871$ & $UVW2$ & 5.5  & 2.6 & $>21.7$ & --   & 230  & {\it Swift} \\
    60528.220 & $\phantom{-}9.874$ & $UVM2$ & 3.1  & 2.5 & $>21.7$ & --   & 230  & {\it Swift} \\
    
\end{longtable}

\section{\texttt{Bagpipes} galaxy analysis}
\label{app:host}

The total stellar mass formed is allowed to vary over $\log_{10}(M_\star/M_\odot)\in\mathcal{U}[5,13]$, and the stellar metallicity over $Z/Z_\odot\in\mathcal{U}[0.01,1.55]$ with a uniform logarithmic prior. Dust attenuation is described by the flexible \citet{Salim2018} law, with the $V$-band attenuation free over $A_V\in\mathcal{U}[0,5]$\,mag, a variable deviation from the \citet{Calzetti2000} slope ($\delta\in\mathcal{U}[-0.7,0.3]$) and a variable $2175$\,\AA\ bump strength ($B\in\mathcal{U}[0,5]$). Following the two-component prescription of \citet{Charlot2000}, stars within their birth clouds (age $<t_{\rm BC}=10$\,Myr) receive twice the attenuation of the diffuse interstellar medium ($\eta=2$). The ionization parameter of the nebular component is varied over $\log U\in\mathcal{U}[-4,-2]$. We further include thermal dust emission using the \citet{Draine2007} models, with free PAH mass fraction, minimum radiation-field intensity, and the fraction of dust exposed to that minimum field.

\begin{figure*}
    \centering
    \includegraphics[width=0.8\linewidth]{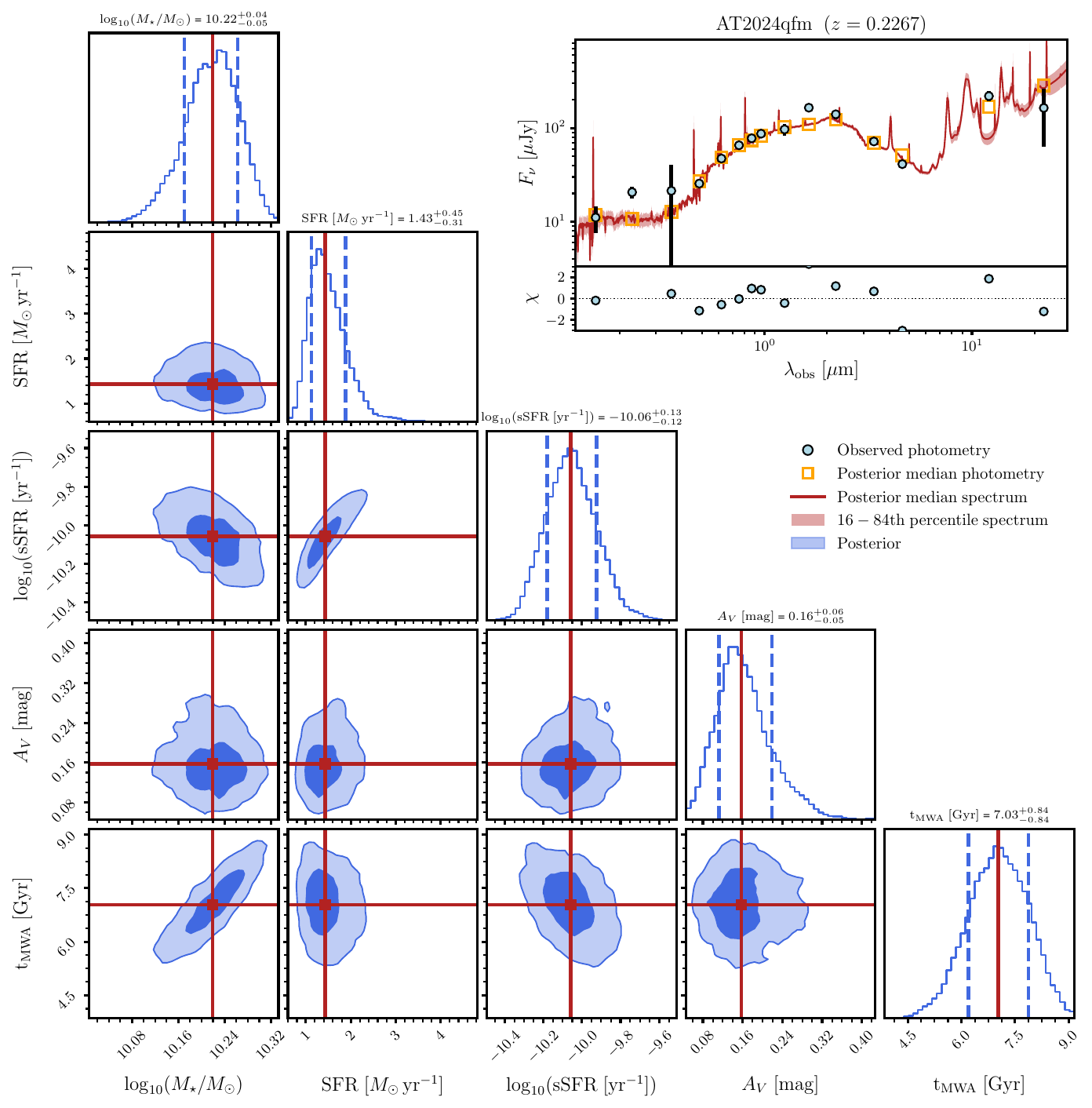}
    \caption{
        Results of our \texttt{Bagpipes} fit to the broadband SED of the host galaxy of \qfm. The main panel shows the marginalised posterior distributions for the key stellar population parameters: stellar mass ($M_*$\,[\msun]), star formation rate (SFR\,[$M_\odot \, \text{yr}^{-1}$]) and specific SFR (sSFR\,[$\text{yr}^{-1}$]), both averaged over the past 100\,Myr, dust attenuation ($A_V$), and mass-weighted age ($t_\text{MWA}$\,[Gyr]). Contours enclose to $1 \sigma$ and $2 \sigma$ credible regions, and the titles quote the posterior median with $1 \sigma$ ($16$th--$84$th percentile) uncertainties; dashed lines mark these percentiles. The inset shows the observed photometry (blue points, with $1\sigma$ error bars), the posterior-median model photometry (orange squares), and the posterior-median model spectrum (red curve) with its $16$th--$84$th percentile range shaded. Residuals relative to the median model are shown in the lower sub-panel.
    }
    \label{fig:hostCorner}
\end{figure*}

\bsp
\label{lastpage}
\end{document}